\documentclass[10pt,preprint]{aastex}

\shorttitle{Dual-beam polarimetry of A Protoplanetary Disk around AB Aurigae}
\shortauthors{Hashimoto et al.}

\begin{document}
  \title{Direct Imaging of Fine Structures in Giant Planet Forming Regions of \\ the Protoplanetary Disk around AB Aurigae\footnote{
    Based on data collected at the Subaru Telescope, which is operated by the National Astronomical Observatory of Japan.}}

  \author{
    J. Hashimoto\altaffilmark{1}, M. Tamura\altaffilmark{1}, T. Muto\altaffilmark{2},
    T. Kudo\altaffilmark{3}, M. Fukagawa\altaffilmark{4}, T. Fukue\altaffilmark{1},
    M. Goto\altaffilmark{5}, C. A. Grady\altaffilmark{6}, T. Henning\altaffilmark{5}, K. Hodapp\altaffilmark{7},
    M. Honda\altaffilmark{8}, S. Inutsuka\altaffilmark{9}, E. Kokubo\altaffilmark{1},
    G. Knapp\altaffilmark{10}, M. W. McElwain\altaffilmark{10}, M. Momose\altaffilmark{11},
    N. Ohashi\altaffilmark{12}, Y. K. Okamoto\altaffilmark{11}, M. Takami\altaffilmark{12},
    E. L. Turner\altaffilmark{10,13}, J. Wisniewski\altaffilmark{14}, M. Janson\altaffilmark{15},
    L. Abe\altaffilmark{16}, W. Brandner\altaffilmark{5}, J. Carson\altaffilmark{5,17},
    S. Egner\altaffilmark{3}, M. Feldt\altaffilmark{5}, T. Golota\altaffilmark{3},
    O. Guyon\altaffilmark{3}, Y. Hayano\altaffilmark{3},
    M. Hayashi\altaffilmark{3}, S. Hayashi\altaffilmark{3}, M. Ishii\altaffilmark{3},
    R. Kandori\altaffilmark{1}, N. Kusakabe\altaffilmark{1}, T. Matsuo\altaffilmark{1},
    S. Mayama\altaffilmark{18}, S. Miyama \altaffilmark{1}, J.-I. Morino\altaffilmark{1},
    A. Moro-Martin\altaffilmark{19,10}, T. Nishimura\altaffilmark{3}, T.-S. Pyo\altaffilmark{3},
    H. Suto\altaffilmark{1}, R. Suzuki\altaffilmark{20}, N. Takato\altaffilmark{3}, 
    H. Terada\altaffilmark{3}, C. Thalmann\altaffilmark{5}, D. Tomono\altaffilmark{3},
    M. Watanabe\altaffilmark{21}, T. Yamada\altaffilmark{22}, H. Takami\altaffilmark{3},
    T. Usuda\altaffilmark{3}
    }
  
  \altaffiltext{1}{National Astronomical Observatory of Japan, 2-21-1 Osawa, Mitaka, Tokyo 181-8588, Japan; jun.hashimoto@nao.ac.jp, motohide.tamura@nao.ac.jp}
  \altaffiltext{2}{Tokyo Institute of Technology, 2-12-1 Ookayama, Meguro, Tokyo 152-8551, Japan}
  \altaffiltext{3}{Subaru Telescope, 650 North A'ohoku Place, Hilo, HI 96720, USA}
  \altaffiltext{4}{Osaka University, 1-1, Machikaneyama, Toyonaka, Osaka 560-0043, Japan}
  \altaffiltext{5}{Max Planck Institute for Astronomy, Heidelberg, Germany}
  \altaffiltext{6}{Eureka Scientific and Goddard Space Flight Center, Greenbelt, MD 20771, USA}
  \altaffiltext{7}{University of Hawaii,640 North A'ohoku Place, Hilo, HI 96720, USA}
  \altaffiltext{8}{Kanagawa University, 2946 Tsuchiya, Hiratsuka, Kanagawa 259-1293, Japan}
  \altaffiltext{9}{Nagoya University, Furo-cho, Chikusa-ku, Nagoya, Aichi 464-8602, Japan}
  \altaffiltext{10}{Department of Astrophysical Sciences, Princeton University, NJ 08544, USA}
  \altaffiltext{11}{Ibaraki University, 2-1-1 Bunkyo, Mito, Ibaraki 310-8512, Japan}
  \altaffiltext{12}{Institute of Astronomy and Astrophysics, Academia Sinica, P.O. Box 23-141, Taipei 10617, Taiwan}
  \altaffiltext{13}{Institute for the Physics and Mathematics of the Universe, The University of Tokyo, Kashiwa 227-8568, Japan}
  \altaffiltext{14}{University of Washington, Seattle, Washington, USA}
  \altaffiltext{15}{University of Toronto, Toronto, Canada}
  \altaffiltext{16}{Laboratoire Hippolyte Fizeau, UMR6525, Universite de Nice Sophia-Antipolis, 28, avenue Valrose, 06108 Nice Cedex 02, France}
  \altaffiltext{17}{Department of Physics and Astronomy, College of Charleston, 58 Coming St., Charleston, SC29424, USA}
  \altaffiltext{18}{The Graduate University for Advanced Studies, Shonan International Village, Hayama-cho, Miura-gun, Kanagawa 240-0193, Japan}
  \altaffiltext{19}{Department of Astrophysics, CAB - CSIC/INTA, 28850 Torrej'on de Ardoz, Madrid, Spain}
  \altaffiltext{20}{TMT Observatory Corporation, 1111 South Arroyo Parkway, Pasadena, CA 91105 USA}
  \altaffiltext{21}{Department of Cosmosciences, Hokkaido University, Sapporo 060-0810, Japan}
  \altaffiltext{22}{Astronomical Institute, Tohoku University, Aoba, Sendai 980-8578, Japan}

  \begin{abstract}    
    We report high-resolution 1.6
   $\micron$ polarized intensity ($PI$) 
   images
   of the circumstellar disk around the Herbig Ae star AB Aur
   at a radial distance of 
   22 AU ($0.''15$) up to 554 AU (3.$''$85), 
   which have been 
   obtained by the high-contrast instrument
   HiCIAO  with the dual-beam polarimetry. We revealed 
   complicated and asymmetrical structures in the inner part ($\lesssim$140 AU) of
   the disk, while confirming the previously reported outer
   ($r$ $\gtrsim$200 AU) spiral structure. 
   We have imaged a double ring structure at $\sim$40 and $\sim$100 AU
   and a ring-like gap between the two. 
   We found a significant discrepancy of inclination angles between two
   rings, which may indicate 
   that the disk of AB Aur is warped. Furthermore, we found seven
   dips (the typical size is $\sim$45 AU or less)  
    within two rings as well as three prominent $PI$ peaks at $\sim$40 AU. 
    The observed structures, including a bumpy double ring, a
   ring-like gap, and a warped disk in the innermost regions, provide
   essential information for understanding 
the formation mechanism of
   recently 
detected
 wide-orbit ($r$ $>$20 AU) planets.  
  \end{abstract}

  \section{Introduction}\label{intro}
  Circumstellar disks are usually formed around young stars and are intricately tied to the origin of planets \citep{shu87,beck96}. 
  Giant planets have been considered to form via gas accretion onto rocky cores in such disks \citep[e.g.,][]{poll96}, 
  which can successfully explain ``normal'' giant planets like ours. However, recent direct detections of companions with 
  masses of a few up to a few tens of $M_{{\rm J}}$
 at distances $>$20 AU, beyond what had been thought to be the planet forming zone
  \citep{maro08,kala08,thal09}, pose a challenge for the standard core-accretion scenario where planets are formed {\it in-situ}. 
  Planet migration after formation such as planet-planet scattering
  \citep[e.g.,][]{vera09}, or gravitational instability (GI) scenarios  \citep[e.g.,][]{duri07,inut10}
  may account for such wide-orbit massive planets.
  To test these theories, information on the detailed structures of the
  inner ($r$ $<$50 AU) regions of protoplanetary disks is crucial.  
  
  One of the best candidates to investigate the inner ($r$ $<$50 AU)
  regions is the prototype young intermediate-mass star AB Aur
  which is one of the most intensively studied Herbig Ae stars 
  \citep[$d = 144^{+23}_{-17}$ pc; $M =$ 2.4 $\pm$ 0.2 $M_{\sun}$; an
  age of 4 $\pm$ 1 Myr;][]{anck97,dewa03}.  
  AB Aur is also known to possess a large ($r$ $>$1000 AU) nebula \citep{grad99}. 
  A more compact ($r$ $\sim$450 AU) rotating disk was detected by
  millimeter observations with a disk mass of $\sim$20 $M_{{\rm J}}$ \citep{henn98}. 
  Near-infrared imaging has revealed a spiral structure in the outer
  ($r$ $\gtrsim$200 AU) part of the disk \citep{fuka04}.  
  A more recent study revealed structures at intermediate distances
  from the star (40 AU $<$ $r$ $<$ 300 AU)  
  and  reported a possible point source in a ``dip'' of the $PI$
  image in the inner disk region \citep{oppe08}. On the other hand,  
  a subsequent study failed to confirm the point source  \citep{perr09}.
  However, all these observations were limited either
  by the inner working distance ($r$ $>$40 AU) or the spatial resolution 
  ($\sim$14 AU). In this $Letter$, we present polarized intensity (hereafter $PI$) 
  images with the smallest inner working angle to the central star ($r$ $>$22 AU) and 
  that the highest spatial
  resolution (9 AU) to date to reveal fine structures in the inner ($r$ $<$50 AU) regions.

  \section{Observations \& Data Reduction}\label{obs}
  $H$-band (1.6 $\micron$) linear polarization images
  of AB Aur were obtained with a newly commissioned high-contrast 
  imaging instrument HiCIAO \citep{tamu06} combined with dual-beam polarimetry 
  on the Subaru 8.2m Telescope on 2009 October 31 UT. The observations were conducted under 
  the program SEEDS \citep[Strategic Explorations of Exoplanets and Disks with Subaru;][]{tamu09}. 
  In the polarimetric observation mode, 
  $o$- and $e$-rays are observed simultaneously, each has 10$''$ by 20$''$ field of view
  with a pixel scale of 9.3 mas/pixel,
  and therefore, $P$I is free from subtraction of reference PSFs.
  For the AB Aur observations, we
  used a small circular occulting mask of 0.$''$3 diameter. 
  The polarizations were measured by rotating the half waveplate to four angular positions (in the order of 0$^{\circ}$, 45$^{\circ}$, 22.5$^{\circ}$, 
  and 67.5$^{\circ}$). We obtained seven data sets by repeating the cycle of waveplate rotations taking a 23.7-s exposure per waveplate position each 
  cycle. The adaptive optics system \citep[AO188;][]{haya04} provided a diffraction limited and mostly stable stellar PSF
  with FWHM of 0.$''$06 in the $H$ band. Low quality images were removed prior to
  final production of the images. 
  We used the four best data sets and the total integration time of the $PI$ image was 189.6 s.
  
  The data were reduced in the standard manner of infrared image reduction using IRAF\footnote{
    IRAF is distributed by the National Optical Astronomy Observatories, which are operated
    by the Association of Universities for Research in Astronomy,
    Inc., under cooperative agreement with the National Science Foundation.},
  namely subtracting a dark frame and dividing by a flat frame. 
  The Stokes $Q$ and $U$ parameters were created by subtracting two split images, in the standard approach for differential 
  polarimetry \citep[e.g.,][]{hink09}. 
  Then, we calculated $PI$ as $\sqrt{Q^{2} + U^{2}}$ shown in figures \ref{f1} and \ref{f2}. 
  $PI$ images have been used by many researchers
  \citep[e.g.,][]{clos97,oppe08} as they have a significant advantage
  over the total intensity (hereafter $I$) images due to 
  difficulties with 
  accurate reference PSF subtractions. 
  
  Since our focus is the $PI$ rather than $I$ image of AB Aur, 
  it is first essential 
  to demonstrate that the $PI$ around 
  AB Aur does not produce any artificial structures. For this purpose, 
  the polarization vector pattern, and its  azimuthal dependence is a good indicator of whether 
  $PI$ is  affected by the bright central star. We calculated averaged azimuthal polarization profiles
  from a reference PSF-subtracted $I$ image, 
  using HD 282411 as the PSF reference star.   
  The reference star HD 282411 was observed just after the observations of AB Aur and was observed 
  in the same observation mode as AB Aur. 
In the PSF subtraction process, 
we
searched for the best subtracted images in the various star-PSF
 frames while varying the registration and scaling to minimize residuals. 
  
\begin{figure}
    \epsscale{0.9}
    \plotone{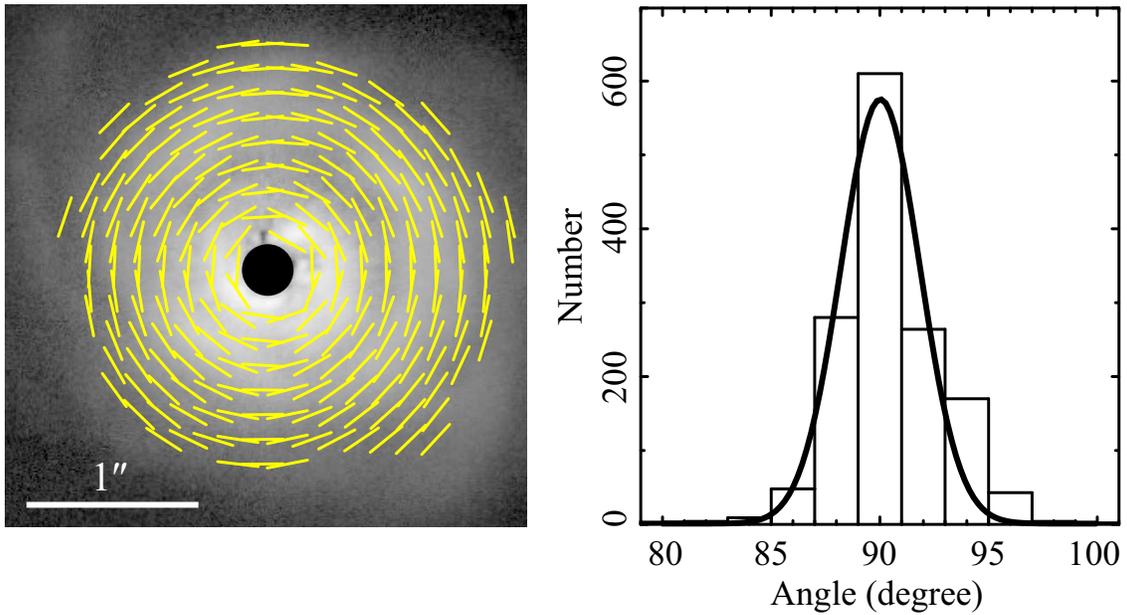}
    \caption{
      Polarization vector pattern of AB Aur.
      Left: Polarization vectors superposed on the $PI$ image of AB Aur in the $H$ band. 
      The plotted vectors are based on 6 $\times$ 6 pixel binning which corresponds to the spatial resolution, 
      and have a polarized intensity larger than 50 $\sigma$. Not all the vectors are plotted and their lengths are not to scale 
      for the purposes of presentation. 
      Right: Histogram of angles between polarization vectors and lines 
      from the mask center to the vector position. As a result of Gaussian fitting, 
      the central position and FWHM are 90.1$^{\circ} \pm$ 0.2$^{\circ}$ and 4.3$^{\circ} \pm$ 0.4$^{\circ}$, respectively. 
    \label{f3}}
  \end{figure}

  Fig.\ref{f3} shows the observed polarization vector 
(the position angle as $0.5 \ {\rm arctan} (U/Q)$)
image of AB Aur 
(see the caption of Fig.\ref{f3} about the construction of the vector image)
and a histogram of the angles 
  between the polarization vectors 
and lines from the stellar position to the vector position. 
  The polarization vector pattern is a good indicator of whether 
  the Stokes $Q$ and $U$ are affected by 
residual speckle noise 
of the bright central star.
  This is because when the Stokes $Q$ and $U$
contain such noise,
  the polarization vectors show either random or parallel alignment.
  As a result of Gaussian fitting
in the histogram,
 we found that the central position and FWHM are 
  90.1$^{\circ} \pm$ 0.2$^{\circ}$ and 4.3$^{\circ} \pm$ 0.4$^{\circ}$, 
respectively. Since the polarization vectors are  
clearly
  centrosymmetric,
  we conclude that the residual speckle noise of AB Aur is quite low and any features identified in our $PI$ images 
  (the ring gap, dips, and peaks) are real.  

  \section{Results \& Discussion}
\begin{figure}
    \epsscale{1}
    \plotone{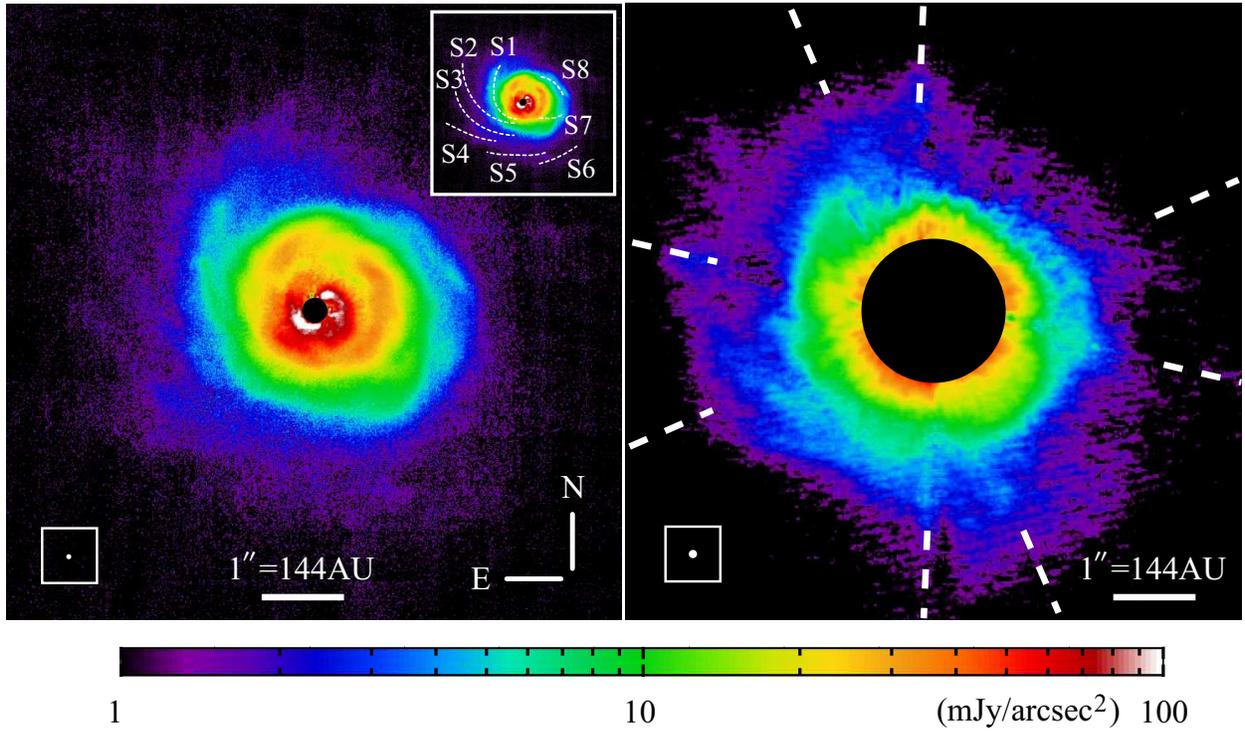}
    \caption{
      Near-infrared ($H$ band) $PI$ and $I$ images of AB Aur.
      Left: The HiCIAO $PI$ image with a coronagraphic occulting mask of 0.$''$3 diameter.
      The multiple spiral structures we refer to as S1 to S8 
      are identified in the top-right inset,
      of which S7 and S8 are newly found.
      Right: The CIAO reference PSF-subtracted $I$ image with a software mask of 1.$''$7 diameter \citep{fuka04}. In contrast to the original image,
      this image shown here is not divided by the square of the radial distance from the star.
      The field of view in both images is 7.$''$5 by 7.$''$5.
      The solid circles in the left-bottom inset in both images represent the spatial resolution of 0.$''$06 and 0.$''$1, respectively.
    \label{f1}}
  \end{figure}

  Fig.\ref{f1} shows  the $PI$ image of AB Aur covering a radial distance of
  22-554 AU ($0.''15$-$3.''85$). 
  Before discussing the inner region, we compare the outer ($r$
  $\gtrsim$140 AU) 
  structures of our $PI$ image and the $I$  
  image of previous observations \citep{fuka04}. The morphology of the
  outer parts of these images is consistent at scales
  larger than the spatial
  resolution,  although one is a $PI$ image and the other is an $I$
  image. These images trace the scattered light of the central star by
  dust grains near the surface of the disk because the disk is optically
  thick at infrared wavelengths. Note, however, that the optically thin
  sub-millimeter continuum radiation also reveals some of the arms; the
  spirals are not simply surface corrugation but also reflect the
  structures near the midplane of the disk \citep{lin06,piet05}. 

\begin{figure}
    \epsscale{0.9}
    \plotone{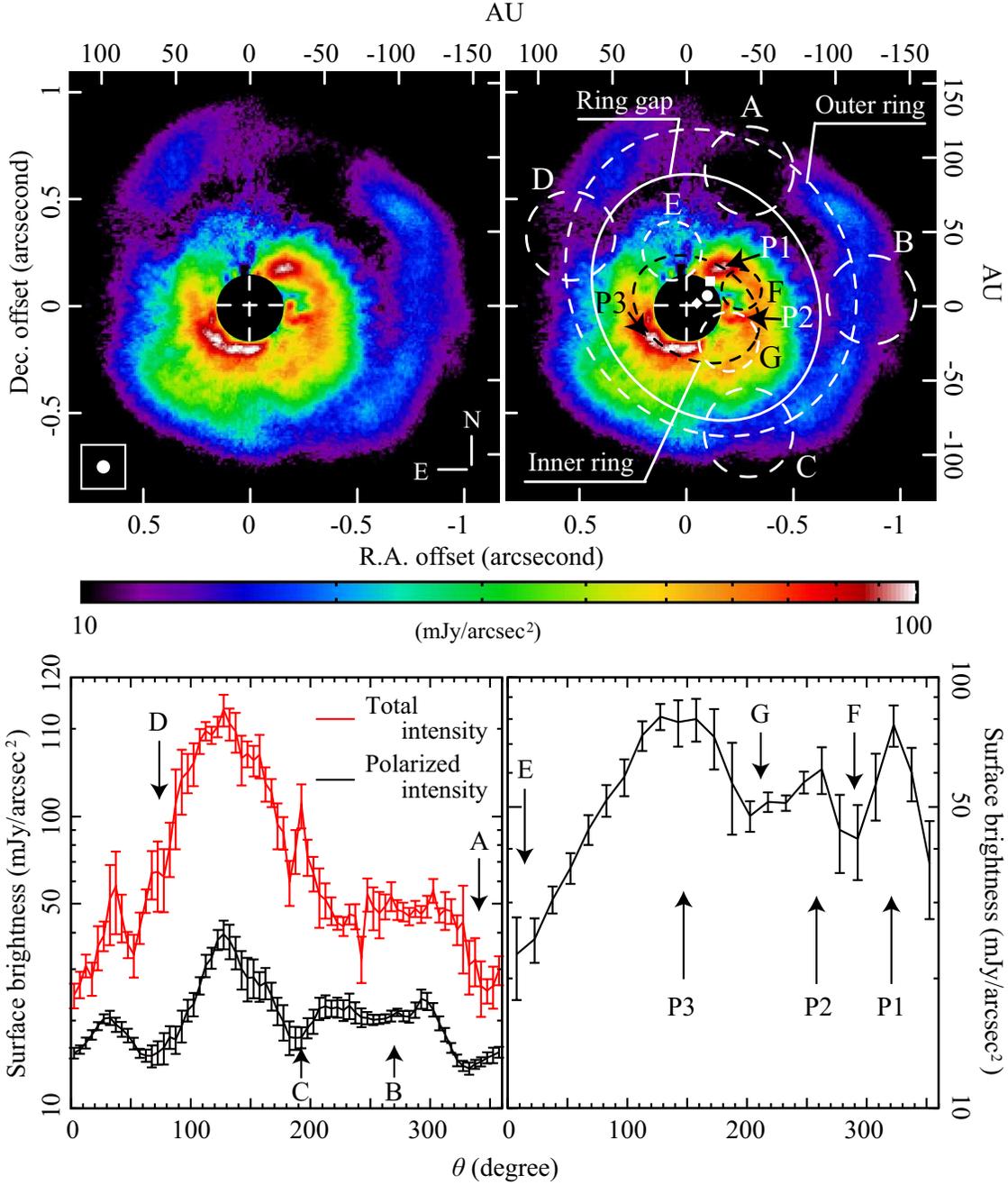}
    \caption{
      Magnified view of the inner $PI$ images of AB Aur and their averaged azimuthal profiles.
      Top: magnified $PI$ image with a coronagraphic occulting mask of 0.$''$3 diameter (left) and the features of the $PI$ image (right).
      Central position (0, 0) is the stellar position. The outer and inner rings are denoted by the 
      dashed ellipsoids. The solid ellipsoid indicates the wide ring gap. The dashed circles (A to G) represent small dips 
      in the two rings. The filled diamond, circle, and square represent the geometric center of the inner ring, ring gap, 
      and outer ring, respectively. The field of view in both images is 2.$''$0 by 2.$''$0.
      The solid circle in the left-bottom inset represents the spatial resolution of 0.$''$06.
      Bottom-left: Averaged azimuthal profiles of the outer ring for the $PI$ (black) and reference PSF-subtracted $I$ (red) images. 
      The profile is averaged every 5$^{\circ}$ in position angle (corresponding to resolution) in the outer ring.
      Bottom-right: Same with the bottom-left image, but for the inner ring with every 15$^{\circ}$ in position angle (corresponding to resolution) 
      in the inner ring.
    \label{f2}}
  \end{figure}

  The real advantage of our new observations is that they allow
  investigation of the inner ($r$ $<$50 AU) structure, comparable to our
  Solar-System.  Fig.\ref{f2} shows a magnified $PI$ image of the inner
  part of AB Aur and the associated azimuthal profiles. The inner regions of the
  disk exhibit complex, non-axisymmetric morphological structures.
  We summarize the significant features revealed in our $PI$ images as follows: 

  (a) We found two ellipse rings, as denoted in Fig.\ref{f2}. Table
  \ref{table1} shows the results of ellipse fitting.  
  The position angles of the minor axes of these two rings are roughly
  consistent with the disk rotation axis inferred from CO emission line
  kinematics \citep[330$^{\circ}$;][]{piet05}. We consider the ellipse
  shape to be due to inclination \citep{laga06};  
  the south-east side of the ring is inclined towards us \citep{fuka04}.  
  The difference of the inclination angles between the inner and outer
  rings suggests that the inner ring might be warped relative to the
  outer ring. 
  In addition, we measured an offset of 0.$''$19 $\pm$ 0.$''$01 between
  the geometric center of the outer ring and the central star.  
  However, since the direction of this offset at 
the far side of the disk with
313.6$^{\circ}$ $\pm$
  5.1$^{\circ}$ is roughly consistent with the disk rotation axis, 
  this offset is most likely due to a geometric effect 
  of the
  flared 
  disk
 \citep[e.g.,][]{fuka04}.
  Our detected double ring structure with a warp has not been reported
  in previous studies
 of AB Aur
 \citep{oppe08,perr09}. 
  
  (b) We found a wide ellipse ring-like gap between the two rings, as
  indicated in Fig.\ref{f2} (hereafter the ring gap).  
  The ellipse fitting results are also shown in table \ref{table1}. This
  ring gap is barely seen in previous $PI$ images \citep{oppe08,perr09}.
  The far-side wall of the ring gap probably corresponds to the
  wall-like structure at 88 AU inferred by mid-infrared observations
  \citep{hond10}. 
  Another example of a gap is HD 100546 \citep{bouw03},
in which
  they inferred that 
this Herbig Be star
 has a disk gap around 10 AU based on the infrared SED. 
  Interestingly, 
they
 also argued that
  AB Aur has a prototype disk without a gap in the inner region.
  
  (c) We found in total seven small dips in the $PI$ within the two
  rings, which we refer to as Dip A to G as shown in Fig.\ref{f2}.  
  The typical size of the dips is $\sim$45 AU or less. The most prominent,
  Dip A at $\sim$100 AU, is consistent with those reported in  
  the $PI$ images of previous studies \citep{oppe08,perr09}. Dip A is
  confirmed at the 3$\sigma$ confidence level in our $I$ image 
  whose averaged azimuthal profile is shown in Figs.\ref{f2}. The
  detection of Dip A in our $I$ image shows that the $PI$ traces  
  the $I$ pattern, and therefore, it is not solely a geometrical
  polarization effect as claimed \citep{perr09}.  
  This detection also suggests that the $PI$ image can be used to
  discuss even in the inner regions where the $I$
  image 
is affected by
 speckle noise. This is especially true when
  discussing the ``local'' structures rather than the ``global'' 
  structures that can be affected by the inclination of the disk.
  In addition, we found three $PI$ peaks in close vicinity to the occulting mask, 
  which we refer to as P1 to P3 in Fig.\ref{f2}. The peaks in the outer rings seen in
  our $PI$ image are not labeled here for simplicity.  
  All the peak structures are extended and 
  thus 
 not
   point sources.

  (d) No point-like sources are detected in Dip A in either the $PI$ or
  $I$ images, as opposed to 
\citet{oppe08}. Our observations are consistent with
\citet{perr09}. 
When we assume that a companion is 100 \% polarized 
in the $PI$ image, which is the faintest case as \citet{oppe08}, 
the upper limits of its mass at
5$\sigma$ (the absolute magnitude of 11.7 at the $H$ band)  
of the photon noise in Dip A are 5 and 6 $M_{{\rm J}}$ for an age of 3
and 5 Myr, respectively \citep{bara03}.  
    These derived upper limits 
    of the masses are consistent with that of 1 $M_{{\rm J}}$ inferred by
    the numerical simulations \citep{jang10}. 
    On the other hand, our upper limits for point sources
    in the dips seen in the inner ring are 7 and
    9 $M_{{\rm J}}$ for these ages due to higher photon noise.  

The structures of AB Aur's inner (22-120 AU) disk surface
  described above indicate that the disk is in an active and
  probably early phase of global evolution, and possibly one or more
  unseen planets are being formed in the disk. 
  
  One possible explanation for the non-axisymmetric structures is GI of
  the disk \citep[e.g.,][]{duri07}. If Toomre's $Q$-parameter 
  (defined as $Q$ = $c_{{\rm s}}\kappa$/$\pi$G$\Sigma$, where 
  $c_{{\rm s}}$, $\kappa$, and $\Sigma$ are the sound speed, epicycle
  frequency,  and surface density, respectively) 
  is of the order of unity, GI occurs and a mode with a small number of
  arms is excited, that is, a pattern of the surface density arises that 
  may resemble what we have observed. However, this GI possibility 
  may be rejected for AB Aur (at present) because optically thin 
  sub-millimeter observations indicate that Toomre's $Q$-parameter is of
  the order of ten \citep{piet05}. 
  It 
  may be 
  noted that the disk mass estimate from sub-mm emission has
large uncertainties arising from the 
  uncertainties in the optical properties of the dust particles.

  The presence of unseen planets in the disk can also result in
  perturbations which extend over the disk scale even in the absence of GI. 
  A low-mass planet in a disk excites a spiral density wave that
  co-rotates with the planet \citep{GT79, TTW02}, 
  while a high-mass planet opens a gap in
  addition to the excitation of a spiral \citep{lin86}, 
  thereby inducing a more significant, globally extended perturbation in
  the disk \citep[see][for the review of disk-planet interaction]{papa07}. 
  The gap opens when the amplitude of the perturbation 
  caused by the embedded planet exceeds the order of unity.
  Since the amplitude of the perturbation scales with $q/h^3$,
  where $q$ is the mass ratio between the planet and the central star and
  $h=H/r$ is the disk aspect ratio \citep{TTW02},
  a crude estimate of the gap-opening mass is $q>h^{3}$.
  For the disk around AB Aur, the temperature of the disk at the
  location of the ring gap ($\sim$80 AU) is 20-30 $K$
  \citep{piet05,lin06}, and therefore, the disk aspect ratio is expected to be
  $\sim$0.1. Therefore, a planet with a mass of only $\sim$1 $M_{{\rm J}}$  
  (consistent with the fact that we detected no point sources) situated
  in the ring gap in the vicinity of Dip A can form a gap at these
  distances. Such a planet cannot be directly seen if it is embedded 
  in the disk equatorial plane; however, its perturbation can induce the
  observed structures such as the ring gap and the largest Dip A and 
  is seen as ``shadows'' \citep{jang10}, while other peaks and small
  dips might be due to small perturbations. Furthermore, a warp 
  in the inner region may be explained by the gravitational perturbation 
  from unseen planets \citep{moui97}. It is also noted that, 
  there is a possibility that the inner ring is intrinsically elliptical
  due to the influence of an unseen gravitating object, which could be
  another indication of the presence of a planet \citep{kley06}.  
  From the present data, it is rather difficult to distinguish the cause of 
  the elliptical shape of the inner ring: either a warped circular ring or
  an intrinsically elliptical ring.  However, we consider that both
  possibility may be accounted for by at least one gravitating object
  embedded in a disk.

  The perturbation caused by an embedded planet
  generally tends to co-rotate with the planet, and therefore the
  deviation of the pattern speed from the local rotation speed would be
  smoking-gun proof of the existence of the planet. The pattern speed of
  the spiral structure is given by 
  \begin{eqnarray*}
    \omega = 0.78 \left( \frac{{\rm M}}{ 2.4 {\rm
		   M}_{\sun}}\right)^{\frac{1}{2}} \left( \frac{r_{{\rm
		   p}}}{80 {\rm AU}} \right)^{-\frac{3}{2}} {\rm
		   [deg/yr]}, 
  \end{eqnarray*}
  where $M$ is the mass of the central star and $r_{{\rm p}}$ is the
  orbital radius of the planet. Such time variability can be observed
  for the next 
  several years. We note that the existence of a planet in the AB Aur
  system, whose age is only 3-5 Myr, may pose a unique constraint 
  on the planet-formation timescale because their formation via gas
  accretion have been considered to take about 10 Myr 
  \citep[e.g.,][]{poll96}. 
  
  Another intriguing explanation for the observed structure is 
  magneto-rotational instability (MRI)
  \citep{balb98}. Although 
  global numerical simulations of MRI is numerically
  challenging, some MRI calculations show that perturbation 
  may extend over the disk \citep{stei02}. 
  It is also shown that MRI drives the disk wind which causes a
  significant perturbation at the disk surface~\citep{SMI10}. 
  In this case, the timescale of variability is on the order of  
  the local rotation timescale, which is longer than that caused by the
  inner unseen planet. 

  In summary, the fine structures including the double ring structure
  with a warp as well as the ring gap detected by our
  observations most likely have an origin in planetary perturbation, 
  but GI or MRI can also be a promising cause of the detailed
  structure. A key future investigation would be the detection of the time
  variability  of the structures, 
  which can provide clues for understanding the formation mechanisms of
  the wide-orbit
 companions
 discovered  
  by direct imaging observations around A stars \citep{maro08,kala08}
  and a G star 
\citep{thal09,jans10}
  as well as a number of physical processes ongoing in the active
  protoplanetary disk.
  
  \bigskip  
  We are grateful to an anonymous referee for providing
  many useful comments leading to an improved paper.
  This work is partly supported by a Grant-in-Aid for Science Research in
  a Priority Area from MEXT and by the Mitsubishi Foundation.
  We also acknowledge support from AST-1008440 (CAG), AST-1009314 (JPW), 
  and a Chretien International Research Grant (JPW).
  
  \clearpage
    
    \begin{table}
      \begin{center}
        \caption{Observational results of AB Aurigae\tablenotemark{a}\label{table1}}
        \begin{tabular}{cccc} 
          \tableline
          \tableline
          & Outer ring  & Ring gap    & Inner ring \\ \tableline
          Diameter of the major axis (AU)               & 210.8 $\pm$ 2.5 & 170.2 $\pm$ 2.0 & 92.0 $\pm$ 6.8 \\
          Diameter of the minor axis (AU)               & 188.1 $\pm$ 2.2 & 149.1 $\pm$ 1.7 & 67.2 $\pm$ 4.2 \\
          Position angle of the major axis ($^{\circ}$)  & 36.6 $\pm$ 6.6  & 36.2 $\pm$ 5.2  & 64.6 $\pm$ 8.9 \\
          Inclination\tablenotemark{b} ($^{\circ}$)      & 26.8 $\pm$ 1.9  & 28.8 $\pm$ 1.7  & 43.1 $\pm$ 6.8 \\
          Geometric center\tablenotemark{c} (mas : mas) & (121 $\pm$ 7 : 127 $\pm$ 8) & (92 $\pm$ 4 : 40 $\pm$ 7) & (54 $\pm$ 17 : 4 $\pm$ 13) \\
          Width\tablenotemark{d} (AU)                   & 29.8 $\pm$ 1.3  & 16.1 $\pm$ 1.6  & 32.1 $\pm$ 2.5 \\ \tableline
        \end{tabular}
        \tablenotetext{a}{
          In the ellipse fitting for the two rings and ring gap, the peak and bottom positions were first directly determined by the averaged 
          radial profile at position angles every 5$^{\circ}$ and 15$^{\circ}$ (corresponding to our spatial resolution) for the two rings and the ring gap, 
          respectively. We then conducted an ellipse fit using an implementation of the nonlinear least-squares Marquardt-Levenberg algorithm with five 
          free parameters of lengths for the major and minor axes, position angle, and central positions.
        }
        \tablenotetext{b}{
          Derived from the ratio of the major and minor axes.
        }
        \tablenotetext{c}{
          Central position (0, 0) is corresponding to the stellar position.
        }
        \tablenotetext{d}{
          Assuming circular structures with given inclinations. 
          The values of the full width at half maximum are derived from Gaussian fitting of the radial profile 
          at the given position angle of the major axis in the north east part.
        }
      \end{center}
    \end{table}
  
  \clearpage

  \clearpage

  \clearpage
  
\end{document}